\documentclass[aps,prl,twocolumn,groupedaddress,amsfonts,showpacs]{revtex4}
\usepackage{graphicx,color}
\usepackage{bm}
\usepackage[latin1]{inputenc}

\begin{document}

\title{Weak localization in ferromagnetic (Ga,Mn)As nanostructures}

\author{D. Neumaier}
\email{daniel.neumaier@physik.uni-regensburg.de}
%\homepage{http://www.physik.uni-regensburg.de/forschung/weiss/}
\author{K. Wagner}
\author{S. Gei{\ss}ler}
\author{U. Wurstbauer}
\author{J. Sadowski}
\author{W. Wegscheider}
\author{D. Weiss}

\affiliation{Institut f\"{u}r Experimentelle und Angewandte Physik,
Universit\"{a}t Regensburg, 93040 Regensburg, Germany}

\date{\today}

\begin{abstract}

We report on the observation of weak localization in arrays of
(Ga,Mn)As nanowires at millikelvin temperatures. The corresponding
phase coherence length $L_{\phi}$ is typically between 100 nm and
200 nm at 20 mK. Strong spin-orbit interaction in the material is
manifested by a weak anti-localization correction around zero
magnetic field.

\end{abstract}


\maketitle

Quantum corrections to the resistance like weak localization are
suppressed by a sufficiently strong perpendicular magnetic field
\emph{B} \cite{Bergmann}. Hence the question arises whether such
effects can be observed in ferromagnets which have an intrinsic
magnetic induction. While few experimental works explored this
problem \cite{Aprili,Dumpich}, a definite experimental answer is
still lacking. Hence, the advent of the new ferromagnetic
semiconductor material (Ga,Mn)As with significantly smaller internal
field compared to conventional ferromagnets offers a new opportunity
to address such questions. Ferromagnetic semiconductors like
(Ga,Mn)As \cite{Ohno} are interesting materials for spintronics as
well, as they combine ferromagnetic properties with the versatility
of semiconductors \cite{Fabian}. The spin $\frac{5}{2}$-Mn-ions on
regular sites of the zinc-blende lattice of the GaAs host  act as
acceptors thus providing both holes and magnetic moments. The
ferromagnetic order between the Mn-ions is mediated by these holes
\cite{Dietl}. By now ferromagnetism in (Ga,Mn)As is well understood,
allowing to predict Curie temperatures \cite{Dietl},
magnetocrystalline anisotropies \cite{Sawicki} as well as the
anisotropic magnetoresistance effect \cite{Baxter}. In this respect
(Ga,Mn)As is one of the best understood ferromagnetic materials at
all \cite{Jungwirth} and hence suitable as a model system to study
quantum corrections to the conductivity.

Interference effects originating from the charge carriers' wave
nature are barely explored and understood in ferromagnets in general
and in (Ga,Mn)As in particular. To this class of effects belong
universal conductance fluctuations (UCF) \cite{Lee}, the
Aharonov-Bohm (AB) effect \cite{Webb}, weak localization (WL)
\cite{Bergmann}, weak anti-localization (WAL) \cite{Bergmann} and
conductivity corrections due to electron-electron interactions (EEI)
\cite{Altshuler}. Recently the existence of AB oscillations in
ferromagnetic rings was predicted theoretically \cite{Tatara} and
subsequently observed in ferromagnetic Fe$_{19}$Ni$_{81}$-
\cite{Kasai} and in (Ga,Mn)As-nanorings \cite{Konni}. In (Ga,Mn)As
the phase coherence length was extracted from UCFs in nanowires
giving typical values  between 90 nm and 300 nm at 20 mK
\cite{Konni,Vila}. This raises the question whether WL corrections -
or WAL effects - can be observed in ferromagnetic (Ga,Mn)As, a
material in which the spin-orbit (SO) interactions for holes in the
valence band is quite strong.

Below we report the observation of WL and WAL in ferromagnetic
(Ga,Mn)As-wires and films thus demonstrating that WL is not
destroyed by the ferromagnets' magnetization. The effect of WL in
disordered electronic systems - investigated intensively in the past
for non-ferromagnetic materials \cite{Bird} - is due to quantum
interference of two  partial waves traveling the same loop type of
path in opposite directions. This leads to an enhanced probability
of backscattering. As an applied perpendicular $B$-field suppresses
the WL the magnetoconductance is positive \cite{Bergmann}. In the
presence of SO interaction the spin part of the wave function needs
to be taken into account. The two partial waves on time-reversed
closed paths experience a spin rotation in opposite direction
causing (partially) destructive interference \cite{Bergmann}. So SO
interactions leads to reduced backscattering and reverses the sign
of the WL, hence called weak anti-localization. A typical signature
of WAL is a double dip in the magnetoconductance trace
\cite{Bergmann}.

For the experiments two wafers having a 42 nm and a 20 nm thick
(Ga,Mn)As layer were used. Both were grown by low-temperature
molecular beam epitaxy deposited on semi-insulating GaAs(001)
\cite{Wegscheider}. The nominal Mn concentration of the 42 nm layer
was 5.5 \%, of the 20 nm layer 5 \%. The Curie temperature $T_C$ of
the as grown layer was 90 K (42 nm) and 55 K (20 nm), respectively.
The samples' remanent magnetization was always in-plane. Some of the
samples were annealed at 200 °C increasing both carrier density and
$T_C$ \cite{Nottingham}. To investigate phase coherent properties
Hall-bar mesas, individual nanowires and arrays of wires were
fabricated employing optical and electron beam lithography. For
nanowire fabrication we used a %Leo Supra 35
scanning electron microscope equipped with a nanonic pattern
generator and subsequent reactive ion etching. Au contacts to the
devices were made by lift-off technique.
%after brief in-situ ion beam
%etching of the surface to remove native oxide.
The characteristic parameters of the samples investigated are listed
in Tab. I.

\begin{table}
\begin{tabular}[b]{|l|p{0.8cm}|p{0.8cm}|p{0.8cm}|p{0.8cm}|p{0.8cm}|} \hline
Sample & 1a & 2 & 2a & 3 & 4 \\\hline
\emph{L} ($\mu$m) & 60 & 7.5 & 7.5 & 7.5 & 0.37 \\
\emph{w} (nm) & 7200 & 42 & 42 & 35 & 35 \\
\emph{t} (nm) & 20 & 42 & 42 & 42 & 42 \\
Number of wires \emph{N} & 1 & 25 & 25 & 12 & 1 \\
%Dimension & 2D & 1D & 1D & 1D & 1D \\
\emph{t}$_{{anneal}}$ at 200 °C (h)& 8.5 & - & 51 & - & - \\
\emph{n} ($10^{26}/\textrm{m}^3$) & 1.7 & 3.8 & 9.3 & 3.8 & 3.8 \\
%D ($10^{-5}m^2/s$) & 2.5 & 7.6 & 11.8 & 7.6 & 7.6 \\
$\rho$ ($10^{-5}\textrm{ }\Omega\textrm{m}$)  & 13 & 3.5& 1.8 & 3.5 & 3.5 \\
$T_C$ (K) & 95 & 90 & 150 & 90 & 90 \\
\hline
\end{tabular}
\label{Daten} \caption{Length \emph{L}, width \emph{w} and thickness
\emph{t} of the samples. Some of the samples were annealed at 200
°C. Resistivity $\rho$ and carrier concentration \emph{n} were taken
at \emph{T} = 300 mK.}

\end{table}

Magnetotransport was measured in a top-loading dilution
refrigerator. To avoid heating, we used a low frequency (19 Hz) and
low current (25 pA to 200 pA) four probe lock-in technique. As we
see no effects of saturation for the different experiments (UCF, WL
and conductivity decrease) at low $T$, we assume that the effective
electron temperature is in equilibrium with lattice and bath
temperature even at 20 mK.

To search for WL effects in (Ga,Mn)As wires we measured the
resistance of \emph{N} parallel wires to suppress UCFs by ensemble
averaging. A corresponding micrograph of sample 2 with 25 wires is
shown in Fig. 1a. The sample's conductance as a function of a
perpendicular \emph{B} field is shown in Fig. 1b. First we start
with a description of the dominant features observed in experiment.
The pronounced conductance maxima around \emph{B} $\sim$ 0 are due
to the anisotropic magnetoresistance (AMR) effect \cite{Baxter}. For
an in-plane magnetization the conductance is higher than for an out
of plane orientation of $M$ \cite{Ohno}. The conductance drops with
the growth of the magnetization's out-of-plane component and
saturates once $M$ is oriented normal to the surface. The positive
slope of the conductance for still higher \emph{B} is due to
increasing magnetic order \cite{Nagaev}. For temperatures larger
than $\sim$ 90 mK the different $G(B)$ traces are shifted but
without noticeable change of shape. The decreasing $G$ for
decreasing $T$ in Fig. 1b stems from the usual low \emph{T} behavior
of the resistance in (Ga,Mn)As which is plotted in Fig. 1c. With
decreasing \emph{T} the resistance rises both for wires (Fig. 1c)
and extended (Ga,Mn)As films (not shown) and is ascribed to EEI.
Similar low \emph{T} behavior has been reported previously for
conventional ferromagnets, too \cite{Dumpich,Ono}. According to
theory \cite{Lee2} the EEI conductivity correction for 1D systems
goes with $-T^{-1/2}$, for 2D systems with  ln(\emph{T}). The
corresponding conductance correction
$\Delta\sigma=\sigma(T)-\sigma(50mK)$ of our sample 2, taken at
\emph{B} = 0 and at \emph{B} = 3 T is plotted in Fig. 1d vs.
$T^{-1/2}$. The resulting straight lines for both \emph{B} values
demonstrate the expected \emph{T} dependence, prove that the
correction is independent of \emph{B} and hence suggest that EEI is
indeed accountable for the conductance decrease at low \emph{T}. For
the 2D sample 1a, $\Delta\sigma$ was best described by a
ln(\emph{T}) dependence (not shown), as expected for EEI in 2D. The
novel features which are in the focus of this letter appear at still
lower temperatures. At about 50 mK two downward cusps at about
$\pm0.4$ T start to become noticeable and have developed to a
prominent feature at 20 mK.

\begin{figure}
\includegraphics[width=8cm]{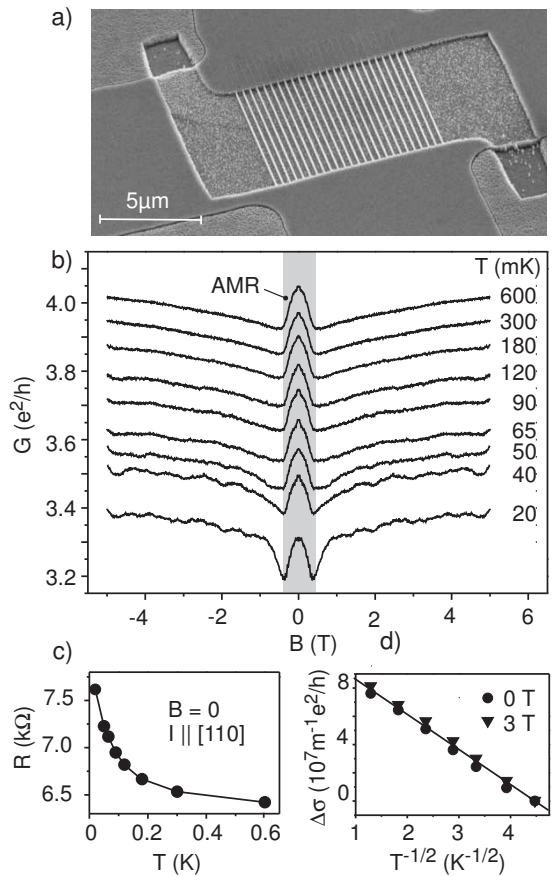}
\caption{(a) Electron micrograph of sample 2 with 25 1D-wires in
parallel. (b) Conductance of sample 2 for different temperatures
measured in a perpendicular magnetic field. To remove the
Hall-conductance in this sample, the antisymmetric part of the
conductance was subtracted \cite{Hall}. The magnetic field range
where the magnetization is rotated from in-plane to out-of-plane is
grey-shaded. (c) Increase of resistance with decreasing temperature.
This increase stems from EEI as proven by the $-T^{-1/2}$ power law
for 1D-systems at \emph{B} = 0 and \emph{B} = 3 T in (d). Here,
$\Delta G $ is taken relative to the conductivity at 50 mK.}
\end{figure}

To separate the peculiar low \emph{T} conductance features from the
"high temperature" background,  $\Delta G=G(\textrm{20 mK})-\alpha
G(\textrm{120 mK})$ of four samples was taken and plotted in Fig. 2.
The factor $\alpha$ takes the $T$ dependence of $G$ into account and
is given by $\alpha=G(\textrm{20 mK})/G(\textrm{120 mK})$. We note,
though, that putting $\alpha=1$ does not change $\Delta G$
qualitatively as the conductance change is only $\sim10$\%. To
compare the different samples, $\Delta G$ was normalized by the
number of parallel wires, \emph{N}. All traces in Fig. 2 show a
characteristic broad conductance minimum for $|B|<$1 T and a local
maximum at $B\sim0$ T. Such $\Delta G(B)$ line shapes are
characteristic for WAL in systems with spin orbit interaction.

\begin{figure}
\includegraphics[width=7cm]{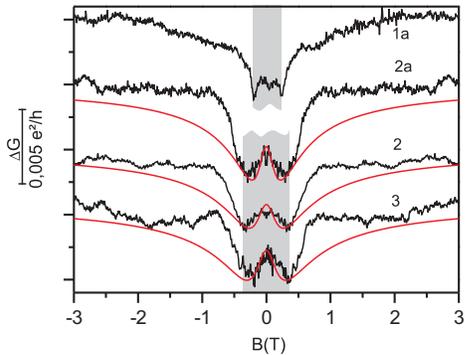}
\caption{WL contribution for three wire and one 2D sample obtained
after subtracting  the 120 mK trace as background conductance. To
compare the different samples the total $\Delta G$ was divided by
the number of parallel wires. In case of the 2D-sample 1a $\Delta G$
was divided by 15 to fit into the graph. Again the grey shaded
\emph{B}-range corresponds to the regime where the samples's
magnetization follows the external field and changes direction. The
red lines are best fits to Eq.(1), discussed in the text. The fit
parameters were $L_{\phi}$ = 190 nm and $L_{SO}$ = 85 nm for sample
2a, $L_{\phi}$ = 150 nm and $L_{SO}$ = 70 nm for sample 2 and
$L_{\phi}$ = 160 nm and $L_{SO}$ = 70 nm for sample 3. Fitting the
2D sample requires a different formalism which is beyond the scope
of the present work.}
\end{figure}

To extract the characteristic lengths from the WL correction we
compare the data of Fig. 2 with existing theory. In Fig. 3a we
particularly compare the WL correction of sample 3, with the
standard expression for WL correction in 1D. Since the width
\emph{w}  and thickness \emph{t} of our wires are smaller than the
phase coherence length $L_\phi$, $w \sim t<L_\phi<<L$ holds and the
1D assumption is justified. The corresponding equation for the
conductance correction reads \cite{Altshuler2,Pierre}:
\begin{eqnarray}
\Delta G=g_s\frac{e^2}{h}\left[\frac{1}{2L}\left(\frac{1}{L_\phi^2}+\frac{1}{3}\frac{w^2}{L_H^2}\right)^{-1/2}\right.\nonumber \\
-\left.\frac{3}{2L}\left(\frac{1}{L_\phi^2}+\frac{4}{3L_{SO}^2}+\frac{1}{3}\frac{w^2}{L_H^2}\right)^{-1/2}\right],
\end{eqnarray}
where $g_s$ is the spin degeneracy. Here, $L_{SO}=\sqrt{D\tau_{SO}}$
is the spin-orbit length that characterizes the strength of spin
orbit coupling, $L_{\phi}=\sqrt{D\tau_{\phi}}$, and
$L_{H}=\sqrt{\hbar/eB}$ is the magnetic length. Eq. (1) is fitted to
the WL data in Fig. 3a for sample 3. As the valence band is spin
split, the holes are highly spin polarized \cite{Braden}. To account
for spin splitting we use either $g_s=1$ (fully spin polarized) or
$g_s=2$ (spin degenerate) as adjustable parameter. While the fit for
$g_s=1$ nicely matches the conductance minima at $\pm400$ mT as well
as the conductance correction $\Delta G$ the fit for $g_s=2$ is less
satisfying. The parameters used for the fit were $L_{\phi}=160$ nm,
$L_{SO}=70$ nm for $g_s=1$ and $L_{\phi}=90$ nm, $L_{SO}=38$ nm for
$g_s=2$ respectively. Also the WL data of the other samples can be
nicely modeled by Eq. (1) and $g_s = 1$; the corresponding fits and
parameters are given in Fig. 2.

\begin{figure}
\includegraphics[width=8cm]{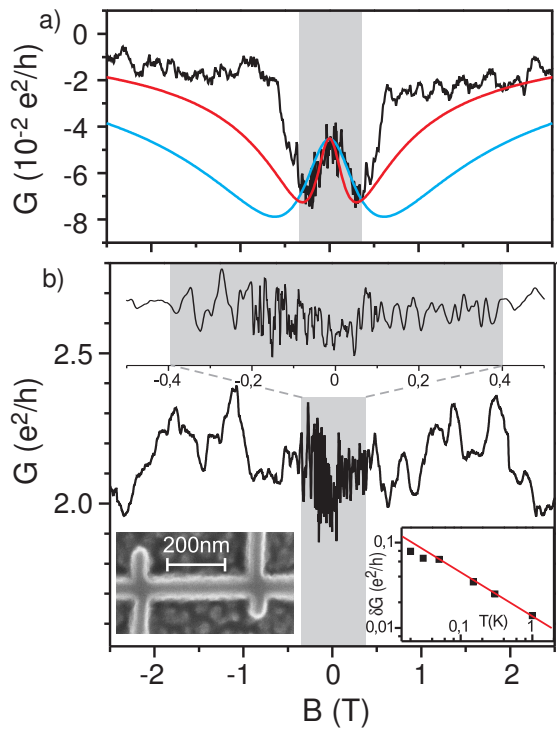}
\caption{(a) WL correction of sample 3 fitted with the standard 1D
WL theory (Eq. 1) for spin degeneracy $g_s=1$ (red) and $g_s=2$
(blue). The parameters used for the fit were $L_\phi=160$ nm,
$L_{SO}=70$ nm for $g_s=1$ and $L_\phi=90$ nm, $L_{SO}=38$ nm for
$g_s=2$, respectively. (b) UCFs measured in an individual 1D-wire
made from the same material (sample 4). An electron micrograph of
the wire is shown in the lower left inset. The grey shaded regime
again corresponds to the magnetic field range where \emph{M} changes
direction. The upper inset shows the low field UCFs in a expanded
magnetic field scale. The temperature dependence of $\delta G$ ,
extracted from the low-field fluctuations, is shown in the lower
right inset.}
\end{figure}

The size of the weak (anti)-localization contribution in Fig. 2 and
Fig. 3a is quite nicely fitted by two parameters, the phase
coherence length $L_\phi$ and the spin orbit length $L_{SO}$ which
is the characteristic length on which spin flip occurs due to SO
interaction. $L_\phi$ can be extracted independently from UCFs
measured on individual 1D-wires \cite{Konni,Vila}, so that
essentially only one free parameter prevails. To study UCFs we
fabricated a single wire, \emph{w} = 35 nm wide and  \emph{L} = 370
nm long, from the same material as sample 2 and 3 (sample 4 in Tab.
1). A corresponding electron micrograph  is shown as lower left
inset in Fig. 3b. $G(B)$ was measured in a perpendicular
\emph{B}-field  from -3 T to 3 T for $T$ between 20 mK and 1 K (for
details see \cite{Konni}). Corresponding data taken at  20 mK show
pronounced, reproducible UCFs, displayed in Fig. 3b. The root mean
square amplitude $\delta G_{rms}=\sqrt{\langle(G-\langle
G\rangle)^2\rangle}$ of these fluctuations is connected with
$L_\phi$ and the wire length \emph{L} by $\delta
G_{rms}\approx(e^2/h)(L_{\phi}/L)^{3/2}f(L_{\phi}/L_{SO})$
\cite{Chandrasekar,TDL}. The function $f(L_{\phi}/L_{SO})$ takes
spin-orbit interaction into account. For $L_{\phi}/L_{SO} \sim 2.3$
we obtain $f(L_{\phi}/L_{SO})\sim 0.53$ \cite{Chandrasekar}.
Extracting $L_\phi$ from  $\delta G_{rms}$, taking only  the
fluctuations between $\pm400$ mT in Fig. 3b into account, results
then in $L_\phi\sim 120$ nm. The temperature dependence of $\delta
G_{rms}$, also taken between $\pm400$ mT is displayed in the lower
right inset of Fig. 3 and shows the characteristic power law
dependence \cite{Konni}. The value of the phase coherence length,
extracted independently from UCFs, is thus in surprisingly good
agreement with the ones used to fit the WL correction. Hence our
analysis suggests that the spin-orbit length $L_{SO}$ ranges between
$\sim$70 nm and $\sim$85 nm in our devices.

While WAL was observed  e.g. in non-magnetic p-type (Al,Ga)As/GaAs
quantum wells \cite{Pedersen} or in (In,Ga)As quantum wells
\cite{Nitta} the observation of WAL-signature in ferromagnetic
(Ga,Mn)As comes as a surprise. A recent theory suggests the
processes, leading to WAL in nonmagnetic systems, to be totally
suppressed in ferromagnets \cite{Dugaev}. In this theory, both the
effect of SO scattering from defects as well as the presence of the
Bychkov-Rashba term was taken into account. The suppression of WAL
in ferromagnets is due to the strong magnetic polarization which
excludes contributions from the so-called singlet Cooperon diagrams,
responsible for anti-localization. As a consequence, the quantum
correction to $G$ is expected to be exclusively negative in
ferromagnets, leading to positive magnetoconductance. This clearly
contradicts our experimental observation.

While the fits in Figs. 2 and 3a are in good agreement with
experiment for $|B|<\textrm{400 mT}$ the concordance at larger $B$
is less perfect. The WL/WAL correction is, as a function of
increasing \emph{B}, more abruptly suppressed than expected from
theory. There is a striking correlation with the magnetic field
dependence of the AMR effect. The magnetic field region where the
AMR occurs is highlighted by grey shading in Fig 1b, 2, 3a and 3b.
Within this \emph{B}-field range the magnetization is rotated from
in-plane to out-of-plane. Once the magnetization is out-of-plane the
WL correction drops quickly. In the same $B$-field range the
fluctuations of an individual wire show a reduced correlation field
$B_C$. Corresponding data are displayed in Fig. 3b, magnified in the
upper inset. Similar behavior was observed in previous experiments
on samples with in-plane easy axis \cite{Konni,Vila} and ad hoc
ascribed to the formation of domain walls in \cite{Vila}. Though we
can not exclude such a scenario we note that $B_C$ is not a well
defined quantity in the regime where the (magnetic) configuration
changes.

The observation of WAL, contrary to theoretical expectation, the
abrupt suppression of the WL correction once the magnetization is
saturated as well as the anomalous $B_C$ in the low \emph{B}-regime
suggest that some important ingredients are still missing to
describe interference phenomena in (Ga,Mn)As. This is not too
surprising as neither the field dependent change of the
magnetization direction nor the $\frac{3}{2}$-spin of the involved
hole states was taken into account. Especially the latter could add
a number of additional interference diagrams not yet treated
theoretically.

In summary we have shown that quantum inference effects strongly
affect the low temperature conductance of ferromagnetic (Ga,Mn)As.
Electron-electron interaction was identified as origin of the
decreasing zero-field conductivity. By resolving a clear weak
localization signature we demonstrate that interference due to
scattering on time reversed paths can exist also in ferromagnetic
materials with internal magnetic induction. The corresponding  phase
coherence length $L_\phi$ in our material, defining the maximum
enclosed area, is between 100 nm and 200 nm at 20 mK and agree with
the values extracted from UCFs. The strong spin-orbit interaction in
(Ga,Mn)As is manifested by a weak anti-localization contribution at
low \emph{B}.

Acknowledgements: We thank K. Richter, I. Adagideli and A. Geim for
valuable discussions. Financial support by the Deutsche
Forschungsgemeinschaft (DFG) via SFB 689 is gratefully acknowledged.

%\begin{references}

% \end{multicols}


\begin{thebibliography}{00}
\bibitem{Hall} In sample 2 and 2a  the two voltage leads were
fabricated on opposite sites. So, also the Hall conductance was
measured. To remove the Hall conductance from the data only the
symmetric part of the trace was taken. As this leads to the same
result than seen on sample 1a and 3, where the voltage leads are on
the same side, physics is not changed by this procedure.
\bibitem{TDL} At 20 mK $L_\phi<L_T=\sqrt{\hbar D/k_BT}\sim$ 170 nm
holds and the thermal lenght $L_T$ needs not be taken into account.
\end{thebibliography}
\end{document}